\documentstyle[preprint,revtex]{aps}
\begin{document}
\draft
\preprint{IFP-437-UNC}
\preprint{TRI-PP-92-96}
\preprint{VAND-TH-92-13}
\preprint{NUHEP-TH-92-20}
\preprint{October,1992}
\vspace{-0.5cm}
\setlength{\baselineskip}{5ex}

\begin{title}
\begin{center}
Searching for dileptons in $Z$ decay
\end{center}
\end{title}

\author{P.H. Frampton, D. Ng{$^a$}}
\begin{instit}
\begin{center}
Institute of Field Physics, Department of Physics and Astronomy,\\
University of North Carolina, Chapel Hill, NC  27599
\end{center}
\end{instit}
\author{T.W. Kephart}
\begin{instit}
\begin{center}
Department of Physics and Astronomy, Vanderbilt University, Nashville, TN 37235
\end{center}
\end{instit}
\author{T.C. Yuan{$^b$}}
\begin{instit}
\begin{center}
Department of Physics, Northwestern University, Evanston, IL  60208
\end{center}
\end{instit}
%
%
\begin{abstract}
The lepton family number violation $Z$ boson decay, $Z \to
e^-e^-\mu^+\mu^+$ or $e^+e^+\mu^-\mu^-$, mediated by a dilepton, e.g.
from an SU(15) theory, is calculated.
The branching ratio of such exotic decay for
allowed dilepton masses is found to be smaller than 10$^{-10}$.
\end{abstract}
Dilepton gauge bosons with lepton number $L = \pm 2$ occur naturally in certain
unified theories as well as in certain non-unified chiral extensions of the
standard model.  For example in SU(15) grand unified theory\cite{1} the
dileptons (generically named $Y$) have been shown to
be heavier than $M_Y \geq 230$ GeV by a study of
polarized muon decay\cite{2}.  In the recent proposal\cite{3} of a three-family
anomaly-free chiral dilepton model based on the group SU(3) $\times$ SU(3)
$\times$ U(1) the dilepton mass satisfied a similar lower bound on $Y$.
In the present
article we shall consider the entire mass range above 100 GeV although it
should be borne in mind that the mass region 100 GeV $< M_Y < 230$ GeV is
allowed only if the dilepton couplings are smaller than predicted by any of the
simplest theoretical models\cite{1,2,3}.

The question addressed in the present article is whether the decay $Z \to
e^-e^-\mu^+\mu^+ + e^+e^+\mu^-\mu^-$ and related decays can put a useful bound
on the dilepton mass with normal dilepton couplings.  Our conclusion will be
that the number of $Z$ decays required is at least 10$^{10}$, a couple of order
of magnitude beyond the accumulation of events likely at, for example, the LEP
collider at CERN.  Thus, this result is a negative one which we report so that
others need not repeat the lengthy calculation.

The interaction Lagrangian\cite{4,5} involving dilepton $Y^{--}$ is given by
\begin{eqnarray}
L(ZY^{--}Y^{++}) &=& \frac{g_2}{\cos \theta_w} \left(-\frac{1}{2} + 2 \sin^2
\theta_w\right) \nonumber\\
&& \mbox{} \times \left [iZ_{\mu\nu}Y^{--\mu}Y^{++\nu}
+ iZ_\mu Y^{--}_\nu Y^{++\mu\nu} + iZ_\nu Y^{--\mu\nu}Y^{++}_\mu
\right] \, ,
\end{eqnarray}
and
\begin{eqnarray}
L(Y^{++}l^-l^-) = -\left(\frac{1}{2}\right)\frac{g_{3l}}{\sqrt{2}} Y^{++}
l^TC\gamma^\mu\gamma_5l\, ,
\end{eqnarray}
where $X_{\mu\nu} = \partial_\mu X_\nu - \partial_\nu X_\mu$,
for $X=Z$ and $Y$.

The five Feynman diagrams contributing to the $Z$ decay $Z \to e^-e^-\mu^+
\mu^+$, which violates the lepton family numbers, are depicted in Fig. 1.

The amplitudes of the Feynman diagrams in Fig. 1 will not be given explicitly.
The branching ratio of the decay $Z \to  e^-e^-\mu^+\mu^+ + e^+e^+\mu^-\mu^-$
is calculated and depicted in Fig. 2 as a function of dilepton mass.  For the
dilepton mass $M_Y > 230$ GeV\cite{2}, the branching ratio is smaller than
10$^{-10}$ when all different lepton families are included.  Therefore, one
still needs greater than 10$^{10}$ samples of $Z$ boson in order to seek this
decay mode.  This is, however, 3 orders of magnitude greater than the
accumulated $Z$ boson at LEP.

Finally, if there exists a $Z^\prime$ which is normally heavier than $Y$ in the
extended standard model, such as SU(15), the $Z^\prime$ decay into two
dileptons will provide a spectacular signature in hadron colliders\cite{6}.

\acknowledgements
The work of P.H.F. was supported in part by the U.S. Department of Energy under
Grant No. DE-FG05-85ER-40219.  The work of D. Ng was supported in part by the
Natural Sciences and Engineering Research Council of Canada and the U.S.
Department of Energy under Grant No. DE-FG05-85ER-40219.
The work of T.W.K. was supported in part by the U.S. Department of
Energy under Grant No. DE-FG05-85ER-40226.  The work of T.C.Y. was
supported in part by the U.S. Department of Energy under Grant
No. DE-FG02-91ER-40684.  Three of us (P.H.F., T.W.K. and T.C.Y.) thank
the Aspen Center for Physics for hospitality while this work was in
progress.



\newpage
\begin{center}
{\bf FIGURE CAPTIONS}
\end{center}

\noindent
Fig. 1.  Feynman diagrams contributing to the decay $Z^o \to e^+e^+\mu^-\mu^-$.

\noindent
Fig. 2.  Branching Ratio for $Z \to e^-e^-\mu^+\mu^+ + e^+e^+\mu^-\mu^-$ as a
function of dilepton mass.

\vfill\eject
{}~
\thispagestyle{empty}
\vfill\eject
\input{psfig}
\centerline{\psfig{figure=zdecay.fig,height=7in}}
\vspace{0.5cm}
\centerline{Figure 2}
\thispagestyle{empty}

\end{document}